\documentclass{IEEEcsmag}
\usepackage[colorlinks,urlcolor=blue,linkcolor=blue,citecolor=blue]{hyperref}
\expandafter\def\expandafter\UrlBreaks\expandafter{\UrlBreaks\do\/\do\*\do\-\do\~\do\'\do\"\do\-}
\usepackage{upmath,color}

\usepackage[table]{xcolor} \usepackage{xspace}

\usepackage[disable]{todonotes}
\usepackage{tikz}
\usetikzlibrary{fit, spy, positioning, calc, arrows, arrows.meta,
  decorations.pathreplacing, decorations.markings, decorations.text, shapes, tikzmark, matrix}

\usepackage{multirow}

\def\checkmark{\tikz\draw[scale=0.4,fill=green!60!black, draw=green!60!black](0,.35) -- (.25,0) -- (1,.7) -- (.25,.15) -- cycle ;}
\def\cross{\tikz\draw[scale=0.4,fill=red, draw=red](.25,0) -- (1,.7) -- (.25,.15) -- cycle (0.75,0.1) -- (0.77,0.1)  -- (0.4,0.7) -- cycle;}
\newcommand{\good}[1]{{\color{green!60!black}{#1}}}
\newcommand{\bad}[1]{{\color{red}{#1}}}

\usepackage{tcolorbox}
\newcommand{\codebackcolor}{gray!10!white}
\definecolor{commentsColor}{rgb}{0.0, 0.388215, 0.488215}
\definecolor{keywordColor}{rgb}{0.000000, 0.000000, 0.635294}
\definecolor{stringColor}{rgb}{0.558215, 0.000000, 0.135316}
\definecolor{keywordCommentColor}{rgb}{0.090000, 0.55, 0.20}

\usepackage{listings}
\tcbuselibrary{listings, skins, raster} 

\lstdefinestyle{cstyle}{%
        language=C,
        basicstyle = \ttfamily\small,
        mathescape = true,
        morekeywords = {u32,u8,u16,bool,int,i8,i16},
        keywordstyle=\color{keywordColor}\bfseries,
        numbers=left,
        xleftmargin=15pt, %
        identifierstyle=\color{black}\ttfamily,
        commentstyle=\itshape\ttfamily\textcolor{commentsColor},
        morekeywords = [2]{pub},
        keywordstyle = [2]\color{green!40!black}\ttfamily\bfseries,
        morekeywords = [3]{key,secret,},       
        keywordstyle = [3]\color{stringColor}\ttfamily\bfseries,
        morekeywords = [4]{mask},
        keywordstyle = [4]\color{brown!40!black}\ttfamily\bfseries,
}

\lstdefinestyle{mipsstyle}{%
        comment = [l]{\#},
        escapeinside={!!},
        frame=none,
        keywordstyle=\color{keywordColor}\bfseries\ttfamily\em,
        identifierstyle=\color{black}\ttfamily,
        commentstyle=\itshape\ttfamily\textcolor{commentsColor},
        stringstyle=\color{Mahogany}\ttfamily,
        basicstyle=\ttfamily\small,
        keywords={, xor, xori, andi, b, lbu, lui, addu, lw, sw, add, addi,
        addiu, jr, jal, nop, move, slti, beqz, blez, mul, },
        morekeywords = [2]{\$zero,\$sp,\$ra,\$a0,\$a1,\$a2,\$a3,\$v0,\$v1,
        \$t0,\$t1,\$t2,\$t3,\$t4,\$t5,\$t6,\$t7,\$t8,
        \$t9,\$s0,\$s1,\$s2,\$s3,\$s4,\$s5, \$fp,\$gp},
        keywordstyle = [2]\color{keywordCommentColor}\bfseries, 
        numbers=left,
        xleftmargin=.08\textwidth,
}

\lstdefinestyle{spstyle}{%
        language=C,
        basicstyle = \ttfamily\normalsize,
        mathescape = true,
        morekeywords = {u32,u8,u16},
        keywordstyle=\color{keywordColor}\bfseries,
        numbers=left,
        xleftmargin=15pt, %
        identifierstyle=\color{black}\ttfamily,
        commentstyle=\itshape\ttfamily\textcolor{commentsColor},
        morekeywords = [2]{Public},
        keywordstyle = [2]\color{green!40!black}\ttfamily\bfseries,
        morekeywords = [3]{Secret},       
        keywordstyle = [3]\color{stringColor}\ttfamily\bfseries,
        morekeywords = [4]{Random},
        keywordstyle = [4]\color{brown!40!black}\ttfamily\bfseries,
}

\usepackage[numbers]{natbib}

\usepackage{subfig}

\jvol{XX}
\jnum{XX}
\paper{8}
\jmonth{Month}
\jname{Publication Name}
\jtitle{Publication Title}
\pubyear{2021}

\begin{document}

\sptitle{}

\title{Protecting Cryptographic Libraries against Side-Channel and
Code-Reuse Attacks}

\author{Rodothea Myrsini Tsoupidi, rtsoupidi@acm.org}
\affil{Independent Researcher$^*$, Stockholm, Sweden}

\author{Elena Troubitsyna, elenatro@kth.se}
\affil{KTH Royal Institute of Technology, Stockholm, Sweden}

\author{Panos Papadimitratos, ppapdim@kth.se}
\affil{KTH Royal Institute of Technology, Stockholm, Sweden}

\markboth{THEME/FEATURE/DEPARTMENT}{THEME/FEATURE/DEPARTMENT}

\begin{abstract}\looseness-1%
Cryptographic libraries, an essential part of cybersecurity, are shown
to be susceptible to different types of attacks, including
side-channel and memory-corruption attacks.
In this article, we examine popular cryptographic libraries in terms
of the security measures they implement, pinpoint security
vulnerabilities, and suggest security improvements in their
development process.\\[5pt]

\textbf{keywords:} cryptographic libraries, code-reuse attacks,
side-channel attacks, compiler-based security
\end{abstract}

\maketitle


\chapteri{C}ryptographic algorithms and protocols constitute an
essential part of software security.
Cryptosystems, that is, ensembles of cryptographic algorithms, provide
security services, such as confidentiality, authentication, integrity,
and non-repudiation, based on mathematical problems deemed
computationally intractable.
Cryptographic libraries implement cryptographic algorithms in software
and they offer an essential building block for implementing any
cryptographic protocol.
The importance of cryptographic libraries in security-critical
applications has attracted malicious attackers that aim at exploiting
the target system.
For example, various cyberattacks~\cite{cachebleed,methodologies} have
demonstrated the presence of exploitable vulnerabilities in different
versions of OpenSSL\footnote{OpenSSL: \url{https://www.openssl.org/}},
a popular cryptographic library, and other cryptographic
  libraries~\cite{blessing2021you}.

Side-channel attacks are powerful attacks that often target
cryptographic software.
These attacks allow unauthorized users to extract sensitive
information during the program execution by recording different
metrics, such as the execution time of the program.
Cryptographic algorithms are specially targeted by side-channel
attacks that aim at extracting secret information, in particular
secret or private cryptographic keys.
Revealing these values may allow an adversary to compromise a secure
communication channel.
For example, the adversary may access and/or modify at will
end-to-end encrypted and authenticated messages, or impersonate a user
using their private key to sign emails~\cite{cachebleed}.

  Another, highly exploitable class of software-enabled attacks is
based on memory leaks.
  Memory-buffer vulnerabilities are among the most common security
  vulnerabilities, comprising approximately 20\% of the reported
  vulnerabilities in cryptographic libraries~\cite{blessing2021you}.
Code-reuse attacks are among the most powerful attacks that are based
on a memory corruption vulnerability, such as a buffer overflow, to
alter the program control flow and hijack the target system.
It is worth noting that many popular cryptographic libraries, such as
OpenSSL, libsodium\footnote{libsodium: \url{https://nacl.cr.yp.to/}},
and cryptlib\footnote{cryptlib:
  \url{https://www.cs.auckland.ac.nz/~pgut001/cryptlib/}}, are
implemented in unsafe programming languages, such as C and
C++~\cite{developers_timing}, which allow memory-corruption
vulnerabilities and unstructured control flow.

To deal with these threats, cryptographic-library developers implement
security measures to protect the library implementation.
These mitigations include source-code countermeasures and
compiler-based mitigations.
Note that timing side channel and code-reuse attacks require different
types of countermeasures and applying these mitigations independently
may compromise the security of the system~\cite{secdivcon}.

A recent survey by \citet{developers_timing} studies  mitigation
techniques against timing side-channel attacks in popular
cryptographic libraries.
The study reveals that only 13 out of 27 libraries claim resistance
against timing attacks.
Moreover, the applied mitigation techniques in some libraries cannot
guarantee the absence of timing side-channel vulnerabilities in the
executable code.
The survey focuses on the developers' effort to protect their code and
the use of verification tools in the testing process of the library to
ensure the absence of timing leaks.
Instead, in this work, we focus on the compilation process as the main
mitigation stage.

Compilers play an essential role in the development of cryptographic
libraries.
In particular, cryptographic libraries typically use general-purpose
optimizing compilers, such as GCC\footnote{GCC:
  \url{https://gcc.gnu.org/}} and LLVM\footnote{LLVM:
  \url{https://llvm.org/}}, to compile the source code and generate
efficient code for a wide range of target processors.
However, code optimization and compiler-based security are two
conflicting goals~\cite{gap}, and general-purpose compilers often lean
towards efficiency.
Instead, secure compilation approaches allow for secure-by-design
development of security-critical applications often at the cost of
code efficiency.

In this work, we investigate the current mitigations of eleven popular
cryptographic libraries against code-reuse attacks and side-channel
attacks and discuss different existing compiler approaches that may
improve security in cryptographic libraries.
Many of these secure compilation approaches have drawbacks, such as
supporting a subset of cyberattacks, introducing high overhead, or
providing low maintenance.
As a way to deal with these open challenges, we propose possible
directions toward improving the compilation of popular cryptographic
libraries:

\begin{itemize}
  \item \textbf{General-purpose compiler developers} need to provide
    additional security countermeasures and control over the
    compilation procedure to the programmer;
  \item \textbf{Secure compilation approaches} need to target multiple
    attacker models, consider optimization goals, and improve
    maintenance;
  \item \textbf{Cryptographic library developers} need to adapt secure
    compilation approaches or enable additional protection options in
    their current compilation scheme.
\end{itemize}

In the following sections, we first describe how timing and code-reuse
attacks operate and common mitigations against these attacks.
Subsequently, we present a summary of compiler-based or
post-compilation countermeasures in cryptographic libraries.
Finally, we present a set of secure compilation approaches that may be
used to protect cryptographic libraries and discuss possible
directions for improving the security of cryptographic libraries.

\section{CYBERATTACKS}

Cryptographic libraries are vulnerable to side-channel attacks.
In particular, side-channel attacks measure time, power consumption,
sound, or electromagnetic emissions, to infer security-critical
values, notably cryptographic keys.
Timing side-channel attacks are particularly effective against
cryptographic implementations because they do not require physical
access to the target device.
For example, OpenSSL and PolarSSL (now Mbed TLS)
  implementations had a timing vulnerability, known as lucky thirteen
  vulnerability (CVE-2013-0169).

Furthermore, a cryptographic library is software that may contain
memory vulnerabilities.
More specifically, memory vulnerabilities threaten cryptographic
implementations.
For example, the Heartbleed vulnerability (CVE-2014-0160) in OpenSSL
allowed reading sensitive memory.
Code-reuse attacks constitute powerful attacks that allow full control
of the system using a memory corruption vulnerabilities.
In general, memory vulnerabilities may enable powerful code-reuse
attacks in two ways: 1) initiate the attack using a memory corruption
vulnerability that is present in the library and 2) use code snippets
from the library code base to stitch a code-reuse attack.

\vspace*{-8pt}

\subsection{Timing Side-Channel Attacks}

Cryptographic implementations are typically written in high-level
languages, such as C and C++.
This code is translated to the target processor machine code, which
consists of a set of hardware instructions.
The execution sequence of these instructions and the values they
process affect the observable execution time of the code.
Precise time measurements allow the observer to extract information
about the processed values.
For example, the execution time of arithmetic division may take
shorter time if the divisor is equal to one than otherwise.

Cryptographic libraries implement cryptographic protocols, which
process security-sensitive values that should remain secret, such as
an encryption key.
The implementation of a cryptographic algorithm may lead to
information leakage during execution.
Figure~\ref{lst:cr} shows the implementation of function
\lstinline[style=cstyle]{check_bit}, which may be part of a naive
implementation of authenticating a user-provided password
(\lstinline[style=cstyle]{pub}) compared to a stored password
(\lstinline[style=cstyle]{key}).
Here, \lstinline[style=cstyle]{u8} denotes an unsigned 8-bit value.
Function \lstinline[style=cstyle]{check_bit} takes two values as
input, \lstinline[style=cstyle]{pub} and
\lstinline[style=cstyle]{key}, and compares them.
If the values are equal, the function returns one, otherwise,
the function returns zero.
We assume that one of the input arguments,
\lstinline[style=cstyle]{key}, is a \textit{secret} value, e.g.\ part
of the password, and the other value, \lstinline[style=cstyle]{pub},
is known to a potential attacker.
At line 3, the code compares the two input values and if they are
equal, the program will assign the value of one to variable
\lstinline[style=cstyle]{t}.
When the comparison is false, the code may take less time to run
(assuming no branch calculation delays) because the processor does not
need to perform the assignment at line 3.
This means that depending on the value of
\lstinline[style=cstyle]{key}, the processor will take less or more
time to execute the function.
Therefore, an attacker that is able to measure the execution time of
this function will be able to derive information about the value of
\lstinline[style=cstyle]{key}.
Typical sources of timing leaks are secret-dependent branches
(e.g.\ Figure~\ref{lst:cr}), secret-dependent memory addressing, and
secret-dependent operands in variable-latency instructions
(e.g.\ division).

\vspace*{-8pt}

\subsection{Code-Reuse Attacks}

\begin{figure}

  \subfloat[][\label{lst:cr} Program with secret-dependent branching]{
  \begin{tcolorbox}[colback=\codebackcolor, colframe=\codebackcolor,
      top=-8pt, bottom = -10pt, right = 0pt, left = 0pt]
    \lstinputlisting[style=cstyle]{cr.c}
\end{tcolorbox}
}

  \subfloat[][\label{lst:bo} Program with buffer overflow]{
  \begin{tcolorbox}[colback=\codebackcolor, colframe=\codebackcolor,
      top=-8pt, bottom = -10pt, right = 0pt, left = 0pt] 
     \lstinputlisting[style=cstyle]{pass.c} 
  \end{tcolorbox}
}

\subfloat[][\label{lst:gadgets} A code-reuse gadget in MIPS32]{
  \begin{tcolorbox}[colback=\codebackcolor, colframe=\codebackcolor,
      top=-8pt, bottom = -10pt, right = 0pt, left = 0pt]
    \lstinputlisting[style=mipsstyle]{original.rop} 
  \end{tcolorbox}
  }
  \vspace{10pt}
\caption{Cryptographic-library vulnerabilities}
\end{figure}

Many cryptographic libraries (see Table~\ref{tab:libs}) are
implemented using unsafe languages with unstructured control flow,
such as C and C++.
The main advantages of these languages are high portability and
increased efficiency.
However, these languages allow the introduction of memory
vulnerabilities that enable powerful attacks via control-flow
hijacking, such as code-reuse attacks.

One direct source of code-reuse vulnerabilities in cryptographic
libraries is the presence of memory corruption vulnerabilities in the
library code.
In particular, code-reuse attacks use a memory corruption
vulnerability, such as a buffer overflow, to insert data into the
program memory and alter the control flow of the program.
The execution of the code may lead to arbitrarily complex
attacker-controlled execution~\cite{rop}.
Memory corruption vulnerabilities, such as buffer overflows, appear
when the program allows a user to write to unintended memory
locations.
For example, buffer-overflow vulnerabilities allow an attacker to
overwrite parts of the memory beyond the limits of an array.
Figure~\ref{lst:bo} shows function
\lstinline[style=cstyle]{read_password}, which reads a user-provided
password to compare with a stored or hashed password.
At line 3, the program defines \lstinline[style=cstyle]{user_password},
an array of eight bytes.
At line 4, the program copies the content of the user input in
\lstinline[style=cstyle]{input} to \lstinline[style=cstyle]{user_password}
using function \lstinline[style=cstyle]{strcpy}.
Neither the main program nor function \lstinline[style=cstyle]{strcpy}
checks whether \lstinline[style=cstyle]{input} fits in
\lstinline[style=cstyle]{user_password}.
Instead, \lstinline[style=cstyle]{strcpy} copies the contents of
\lstinline[style=cstyle]{input} to the memory that starts at the
address of \lstinline[style=cstyle]{user_password}, potentially overwriting
data that follow array \lstinline[style=cstyle]{user_password}.
Typically, compilers store locally defined arrays in the stack frame;
thus, in Figure~\ref{lst:bo}, an input of more than eight characters
results in overwriting the
stack.
The stack frame often contains the address to which the function
will return to, and therefore, overwriting this address leads to a
redirection of the control flow.
An attacker may overwrite the return address of the function with an
arbitrary address in the execution code to redirect the program
execution to this arbitrary address.

The second source of code-reuse vulnerabilities in cryptographic
libraries is the compiled code of the library.
When the attacker identifies a memory corruption vulnerability in the
cryptographic library or other parts of the executable program, the
attacker inserts data that redirects the program execution and results
in unwanted program behavior.
The attacker payload consists of carefully selected code snippets,
so-called gadgets, which when stitched together, result in an attack.
In our context, the attacker may use the library code base as a pool
for code-reuse gadgets that enable code-reuse
attacks~\cite{methodologies}.
A gadget is a code snippet that typically ends with a control-flow
instruction, which allows the attacker to direct the execution to the
next gadget.
Figure~\ref{lst:gadgets} shows a code-reuse gadget extracted from the
GNU C Library (\texttt{libc})\footnote{libc:
  \url{https://www.gnu.org/software/libc/}} in a MIPS-based system
using ROPGadget\footnote{ROPGadget:
  \url{http://shell-storm.org/project/ROPgadget/}}, a code-reuse
gadget extraction tool.
The gadget starts with a load operation, that loads an
attacker-controlled value from the stack,
\lstinline[style=mipsstyle]{0x24($sp)}, to the return-address
register, \lstinline[style=mipsstyle]{$ra}.
The second instruction (line 3) loads an attacker-controlled value to
register \lstinline[style=mipsstyle]{$s0}.
This value may be used by the attacker to perform a malicious
operation, for example, to create a system shell.
At line 4, the gadget jumps to the next attacker gadget, where the
control flow continues to the address in register
\lstinline[style=mipsstyle]{$ra}.
The last line\footnote{The MIPS architecture uses \textit{delay
  slots}, which follow a branch instruction but are executed before the
  branch.} updates the stack pointer
\lstinline[style=mipsstyle]{$sp}, which allows the attacker to use a
different part of the stack for the next gadget.

Code-reuse gadgets are present in almost every code implementation.
Therefore, an executable and any dynamic library that is linked to
this executable can provide the attacker with useful building blocks
to construct an attack.

\vspace*{-8pt}

\section{COUNTERMEASURES}
This section describes countermeasures against timing side-channel
attacks and code-reuse attacks, including both manual source-code
mitigations and compiler-enabled countermeasures.

\subsection{Timing Side-Channel Mitigations}

Timing side-channel attacks are based on observations of
secret-dependent timing variations during code execution.
A well-known programming discipline to mitigate timing side channels
is constant-time programming, where the execution time of the program
should not depend on secret values.
This means that the mitigated program does not contain any
secret-dependent branch instructions, memory operations, or
variable-latency instruction operands.
Constant-time programming requires transforming secret-dependent code
to constant-time equivalent.
Often, developers implement this mitigation manually.
Figure~\ref{lst:ctmitig} shows a naive constant-time implementation of
the code in Figure~\ref{lst:cr}~\cite{verifying}.
The idea of this mitigation is that there is a single assignment of
variable \lstinline[style=cstyle]{t} regardless of the value of
\lstinline[style=cstyle]{key}.
The compiler may translate the \lstinline[style=cstyle]{?:} expression
to constant-time code, such as \texttt{cmov} in x86 or \texttt{csel}
in ARM, which is the desirable result.
However, the compiler may translate the expression to an
\lstinline[style=cstyle]{if-else} statement, breaking the
constant-time mitigation~\cite{binsec}.

\begin{figure*}
  \centering
    \begin{tcolorbox}[colback=\codebackcolor, colframe=\codebackcolor,
        top=-15pt, bottom = -20pt, right = 0pt, left = 0pt]
      \subfloat[][\label{lst:ctmitig} Naive constant-time transform]{
        \hspace{10pt}
        \lstinputlisting[style=cstyle, basicstyle = \ttfamily\small]{empty.c}}
      \hfill
      \subfloat[][\label{lst:ctmitig2} Constant-time transform]{
        \lstinputlisting[style=cstyle, numbers = none, lastline=6, basicstyle = \ttfamily\small]{logic.c}
      }
      \hfill
      \subfloat[][\label{lst:bbbmitig} Constant-resource transform]{
        \lstinputlisting[style=cstyle, numbers = none, lastline=6, basicstyle = \ttfamily\small]{copy.c}
      }
    \end{tcolorbox}
  \vspace{20pt}
\caption{\label{lst:figs} Transformations of the code in Figure~\ref{lst:cr}}
\end{figure*}

Figure~\ref{lst:ctmitig2} shows an alternative constant-time
transformation of the code in Figure~\ref{lst:cr}.
First, this implementation (line 3) creates a mask,
\lstinline[style=cstyle]{m}, which depends on the value of
\lstinline[style=cstyle]{key} and has value \texttt{0xFFFFFFFF} (if
the values of \lstinline[style=cstyle]{key} and
\lstinline[style=cstyle]{pub} are equal) or \texttt{0x00000000} (if
the values are not equal).
Then, the code copies the correct value, zero or one, using the
previously generated mask.
This transformation is more difficult for the compiler to revert.
Nonetheless, different compiler versions and optimization flags may
break this mitigation, by converting the code to a branch
statement~\cite{binsec}.

Constant-time implementations are often implemented manually and are,
therefore, error-prone.
Moreover, the code becomes difficult to understand and analyze.
Figure~\ref{lst:bbbmitig} shows an alternative mitigation often
referred to as constant-resource programming~\cite{secdivcon}.
This mitigation adds dead code to ensure that both branch paths take
the same time to execute (line 6).
To ensure the correctness of this mitigation, the compiler needs to be
aware and preserve this mitigation, otherwise, the compiler may remove
the \lstinline[style=cstyle]{else} branch because
\lstinline[style=cstyle]{_t} is an unused variable.
In addition, this mitigation may require an accurate timing model of the target
processor.

\subsection{Code-Reuse Attack Mitigations}

One direction towards mitigating code-reuse attacks is identifying and
mitigating possible memory-corruption vulnerabilities in the code
implementation.
Identifying these vulnerabilities often includes static analysis of
the code to discover possible vulnerabilities and/or add runtime
buffer-overflow checks.
A disadvantage is the inability to guarantee the absence of
vulnerabilities in the program code and linked-library code, which may
enable a code-reuse attack.

\textit{Stack canaries} is a mitigation against buffer overflows on
the stack.
A canary is a randomly generated word at the end of the stack frame
that is set at runtime.
This word is checked before jumping out of the function to detect
possible stack-smashing attempts.
Apart from hindering illegal data insertion in the stack, stack
canaries may also hinder control-flow violations in code-reuse
attacks, such as Return-Oriented Programming (ROP)~\cite{rop}.
ROP attacks use code-reuse gadgets that end with a return instruction.
Mitigating ROP attacks requires introducing stack canaries at the exit
of every function.
Unfortunately, there are ways around stack canaries, such as
brute-force attacks, memory-disclosure attacks~\cite{ssr_android}, or
Jump-Oriented Programming attacks that use gadgets ending with
non-return branches.

Automatic software diversification or randomization is a method to
protect against most types of code-reuse attacks.
The idea is that introducing randomness to the implementation of the
code reduces the probability of a successful attack.
For example, the gadget in Figure~\ref{lst:gadgets} is located at
relative address \texttt{0x00083290} is \texttt{libc}.
However, if we can relocate this gadget, the attacker payload will
need to adjust the payload for this relocation.
The most widely used diversification scheme against code-reuse attacks
is Address Space Layout Randomization (ASLR).
ASLR randomizes dynamically the base address of the executable and the
addresses of the stack, heap, and dynamically loaded libraries.
ASLR is an important mitigation against code-reuse attacks, however,
brute-force attacks may recover the randomized base addresses.
Fine-grained diversification approaches are typically more difficult
to defeat and require more sophisticated attacks~\cite{methodologies}.

Another method to protect against code-reuse attacks is control-flow
integrity~\cite{cfi}.
Control-flow integrity approaches aim to ensure that the control flow
of the program at runtime complies with the intended control flow.
A main disadvantage of control-flow integrity schemes is high overhead
and/or dependency on specialized hardware.

\vspace*{-8pt}

\section{SECURITY IN CRYPTO LIBRARIES}

\begin{table}
  \vspace*{4pt}
  \caption{\label{tab:libs} Cryptographic libraries with enabled
    compilation flags (CF) and post-compilation timing testing (PCT); SP stands for
    \texttt{-fstack-protector-strong}, SBS stands for \texttt{-param
      ssp-buffer-size}, PIE stands for \texttt{-fPIE}, FORT stands for
    the macro \texttt{-DFORTIFY\_SOURCE}, and CTT stands for
    constant-time testing.}
  \rowcolors{3}{}{gray!25}
  \centering \tablefont
  \begin{tabular}{l|cccc|c}
    \multirow{2}{*}{Library} & \multicolumn{4}{c|}{CF} & PCT~\cite{developers_timing} \\\cline{2-6}
     & SP  & SBS & PIE & FORT           & CTT           \\\hline
  BearSSL   &  \checkmark  & &     &            &  \cross           \\
  Botan     &  \checkmark$^*$&     &      &       &  \checkmark       \\
  cryptlib  &  \checkmark  &     &      &  2    &  -                \\ 
  Crypto++  &  \checkmark  &     &      &       &  \cross           \\
  GnuTLS    &  \checkmark  &     &      &       &  \cross           \\
  LibreSSL  &  \checkmark  &     &      &       &  \cross           \\
  Libgcrypt &  \checkmark  &     &      &       &  -                \\
  libsodium & \checkmark$^*$ &  & \checkmark    & &  \cross           \\
  Mbed TLS  & \checkmark  &     &      &       &  \checkmark       \\
  OpenSSL   & \checkmark  &     &      &       &  \cross           \\
  wolfTLS   &   \checkmark  & 1 &     &            &  \cross           \\\hline 
  \multicolumn{6}{l}{$^*$ the flag is \texttt{-fstack-protector} }\\
  \end{tabular}
\end{table}

Popular libraries use countermeasures to protect against timing side
channels and code-reuse attacks.
We selected eleven well-known cryptographic libraries
that are open source, BearSSL\footnote{BearSSL:
  \url{https://bearssl.org/}} 0.6, Botan\footnote{Botan:
  \url{https://botan.randombit.net/}} 2.19.3, cryptlib 3.4.7,
Crypto++\footnote{Crypto++: \url{https://www.cryptopp.com/}} 8.8.0,
GnuTLS\footnote{GnuTLS: \url{https://www.gnutls.org/}} 3.6.16,
LibreSSL\footnote{LibreSSL: \url{https://www.libressl.org/}} 3.8.0,
Libgcrypt\footnote{Libgcrypt:
  \url{https://gnupg.org/software/libgcrypt/}} 1.10.2, libsodium 1.10.18, Mbed
TLS\footnote{Mbed TLS:
  \url{https://www.trustedfirmware.org/projects/mbed-tls/}} 3.4.1, OpenSSL 3.1.2,
and wolfTLS\footnote{wolfTLS: \url{https://www.wolfssl.com/}} 3.13.0.
The first part of Table~\ref{tab:libs} shows the
compilation flags in GCC or LLVM for the default build process of the
library.
The flags mainly activate buffer-overflow checks (FORT), stack
canaries (SP and SBS), and full ASLR (PIE).
In the table, \checkmark denotes the presence of the flag in the
compilation process, whereas an empty box corresponds to the absence
of the flag.
To extract these flags, we compile each library on an Intel%
\textsuperscript{\textregistered}%
Core\texttrademark i7-8550U
processor at 1.80GHz with 16 GB of RAM running Ubuntu 18.04, using
GCC version 7.5.0 and clang version 10.0.1.
The second part of the table marks which of the libraries perform
post-compilation timing testing during the development process, based
on the results by \citet{developers_timing}.

\subsubsection{Timing Mitigations in Cryptographic Libraries:}
\citet{developers_timing} show that many developers are aware of
timing attacks and possible mitigations.
More specifically, many cryptographic libraries support constant-time
cryptographic algorithms and/or re-implement known vulnerable
algorithms, e.g.\ Advanced Encryption Standard, using the
constant-time programming discipline.
However, most cryptographic libraries are not able to provide
guarantees that the compiled binary code is constant time.
Table~\ref{tab:libs} shows that only two libraries perform
constant-time testing as part of their code development
process~\cite{developers_timing}.
These results indicate that most cryptographic libraries do not
guarantee the absence of timing vulnerabilities.

In addition, \citet{developers_timing} mention the following problems:
1) the uncertainty in preserving security properties in popular
compilers and 2) the inability of current compilers to incorporate the
secret/public type system into their current system.
These challenges require in-depth changes to the compiler
infrastructure to take into account side-channel attacks and their
countermeasures.

\subsubsection{Code-Reuse Mitigations in Cryptographic Libraries:}
Several code-reuse attack mitigations are included as optional or
default in GCC and LLVM.

A method to prevent a buffer overflow is to check the buffer size
before starting to copy the data (see Figure~\ref{lst:bo}).
Preprocessor flag \texttt{-DFORTIFY\_SOURCE} in GCC and clang defines
a macro that inserts code around possible buffer-overflow
vulnerabilities, which performs runtime buffer-overflow checks.
In particular, the macro replaces common vulnerable functions, such as
\texttt{strcpy}, with wrapper functions that check whether the size of
the destination buffer is large enough to store
the source data.
In Table~\ref{tab:libs} only cryptlib uses this flag during
compilation.

In GCC and LLVM, flags \texttt{-fstack-protector} and
\texttt{-fstack-protector-strong} add stack canaries to functions that
may contain buffer overflows, namely functions that manipulate a
buffer of at least \texttt{-param ssp-buffer-size} buffer size.
All libraries in Table~\ref{tab:libs} use either of these flags with
the default buffer size, eight bytes, apart from wolfTLS, which uses a
smaller buffer size of one byte.
Flag \texttt{-fstack-protector-all} provides stronger protection as it
adds stack canaries to every function and can protect against ROP
attacks.
Unfortunately, none of the cryptographic libraries in
Table~\ref{tab:libs} uses this flag.

ASLR is default in many operating systems, however, the address of the
executable code is typically not randomized.
In GCC and LLVM, flag \texttt{-fPIE}, enables ASLR for executables.
This flag is only used by libsodium in Table~\ref{tab:libs}.

In summary, Table~\ref{tab:libs} shows that all libraries implement
stack canaries in selected functions, however, the majority of the
libraries do not enable additional compiler mitigations against
code-reuse attacks.

\subsubsection{Cryptographic-Library Security Challenges:}

Based on the work by \citet{developers_timing} and our experiment,
cryptographic-library developers appear to be aware of the importance
of security mitigations in cryptographic code.
In particular, developers use a set of mitigations against timing side
channels and code-reuse attacks.
For mitigating timing side channels, developers depend often on manual
implementations, whereas for mitigating code-reuse threats, developers
use the available compiler options.

Unfortunately, these mitigations are not enough to guarantee security;
the compilation process may remove source-code mitigations, while the
features and options that compilers provide do not guarantee the
mitigation of advanced code-reuse attacks.
For example, stack canaries (used by all libraries) are not enough to
protect against different types of code-reuse attacks, such as
JOP attacks.
Furthermore, cryptographic-library developers do not perform
binary-level testing to ensure that source-code mitigations remain in
the compiler-generated code.
One solution to these challenges is to incorporate security
checks and mitigations during compilation.
\vspace*{-8pt}

\subsection{Position-Independent Execution (PIE)}
The experiment we performed in this section considers the occurrence of
the \textit{-fPIE} during the compilation process.
However, this flag affects the generated static libraries (\textit{.a}) 
and not dynamic libraries (\textit{.so}).
In particular, an executable compiled with \textit{-fPIE}, will be
position independent when dynamically linked (using the
linking flag \textit{-pie}) to a library not compiled with
\textit{-fPIE}, but with \textit{-fPIC}.
Dynamic shared libraries built 
using \textit{-fPIC/-fpic} may be 
loaded at randomized addresses when ASLR is enabled. 
It is also worth mentioning that different operating systems may
provide compilers with different default flags.
For instance, in Fedora 39, the precompiled version of GCC (version
13.3.1-3) does not include any flag to enable PIE by default, whereas
in NixOS 25.11 and different Ubuntu versions, PIE is enabled by
default.

\section{SECURE COMPILATION}

We present four secure compilation approaches integrated into
conventional compilers, in particular, LLVM, a popular and freely
available compiler infrastructure toolchain.
We selected a set of diverse compiler approaches with regards to the
method they use, however, this list is not complete.
Table~\ref{tab:tools} shows information about these LLVM-compatible
compilers, including the attacks they mitigate and the overhead they
introduce (\good{green} denotes advantageous values,
  \bad{red} denotes disadvantageous values).
In the next sections, we describe each of these approaches.
\vspace*{-8pt}

\begin{table*}
  \vspace*{4pt}
  \caption{\label{tab:tools} Compiler tools; LLVM lists the version of
    LLVM the tool is based on; TSC means that the tool provides
    mitigation(s) against timing side-channel attacks; CRA means that
    the tool provides mitigation(s) against code-reuse attacks; Avail
    lists whether the tool is publicly available; ETO$_{a}$ stands
    for Execution-Time Overhead and CTO$_{a}$ stands for
    Compilation-Time Overhead for $a \in \{TA, CR\}$.}  \centering
  \rowcolors{2}{}{gray!25} \tablefont
\begin{tabular}{l|lcccllll}
  Compiler            & LLVM & TSC & CRA & Avail & ETO$_{TA}$ & CTO$_{TA}$ & ETO$_{CR}$ & CTO$_{CR}$ \\\hline
  SecComp~\cite{vu}   & \good{$\sim$11}  & \checkmark & \cross & \cross & \good{low} & \good{low} & - & - \\
  MCR~\cite{multicomp} & \bad{3.8} &\checkmark & \checkmark & \checkmark &  \bad{high} & \good{low} & \good{low} & \good{low} \\
  SecDivCon~\cite{secdivcon} & \bad{3.8} &\checkmark & \checkmark & \checkmark & \good{low} & \bad{high} & \good{low} & \bad{high}\\
  PCFL~\cite{linearization} & \good{13}  & \checkmark & \cross &  \checkmark & \bad{high} & \good{low} & - & - \\\hline
\end{tabular}
\end{table*}

\subsection{Secure Compiler}

The \citet{vu} approach targets multiple attacker models, including
timing side channels, with the compiler based on a recent version of
LLVM and evaluated for x86 systems and ARMv7-M/Thumb-2.
Their paper introduces the notion of \textit{opaque observations},
marking operations the compiler cannot remove or replace, while it can
statically analyze them.
\textit{Opaque observations} are source-code annotations and are
preserved throughout the compilation procedure, forcing the compiler
to preserve constant-time source-code mitigations.
  To ensure that the code mitigation in Figure~\ref{lst:bbbmitig} is
  valid after compilation using the compiler by \citet{vu}, the
  developer may write the following code:
      \begin{tcolorbox}[colback=\codebackcolor, colframe=\codebackcolor,
        top=-8pt, bottom = -10pt, right = 0pt, left = 0pt]
        \lstinputlisting[style=cstyle, numbers = none, showlines=true,
          basicstyle = \ttfamily\small]{copy_sec.c}
    \end{tcolorbox}
  Code annotation, \lstinline[style=cstyle]{// Property: observe(_t)},
  ensures that the compiler will not remove the unused code in the
  \lstinline[style=cstyle]{else} statement in
  Figure~\ref{lst:bbbmitig}.
\citet{vu} provide improved speedup compared to unoptimized LLVM code,
often considered secure, at a small compilation-time overhead.

\citet{vu} do not deal with code-reuse attacks.
However, their approach may use the mitigations that LLVM provides,
such as stack canaries.
Unfortunately, these mitigations may be defeated by advanced attacks,
such as ROP or JOP attacks.
To reduce the effect of these attacks, a developer may combine
\citet{vu} approach with additional mitigations against code-reuse
attacks.
However, this comes at a risk of breaking the security properties of
the tool.
To summarize, the approach by \citet{vu} is a promising solution to
the security/optimization gap and provides a solution that is
compatible with the compilation procedures of modern cryptographic
libraries.
We believe that a similar solution can be adopted by compiler
developers of GCC and LLVM to provide mitigations against more
security threats with limited efficiency overhead.
With regard to code-reuse attacks, \citet{vu} do not provide advanced
mitigations, such as control-flow integrity or fine-grained
randomization.
An important hindrance to adopting this approach is that it requires
some manual annotation, and more importantly, it is not freely
available.

\subsection{Multicompiler}
Multicompiler (MCR)~\cite{multicomp} is an open
source\footnote{MCR: \url{https://github.com/securesystemslab/multicompiler.git}},
program randomization tool, based on LLVM 3.8, targeting x86
architectures.
MCR supports different levels of software randomization with
main focus on mitigating code-reuse attacks~\cite{multicomp}.
MCR takes a random seed as a command-line argument and generates
diverse code given different random seeds.
The tool does not require any additional annotations.

MCR compiles a C program using a modified version of LLVM.
The tool supports stack layout randomization using flag \texttt{-mllvm
  -shuffle-stack-frames}, function randomization with \texttt{-mllvm
  -randomize-function-list}, no-operation insertion with
\texttt{-Xclang -nop-insertion}, hardware register randomization with
\texttt{-mllvm -randomize-machine-registers} and more.
To generate diverse programs in every compilation, MCR uses
  \texttt{-frandom-seed=<seed>}.
For example, to compile a file name password.c, you can write:
  \begin{tcolorbox}[colback=\codebackcolor, colframe=\codebackcolor,
        top=-8pt, bottom = -10pt, right = 0pt, left = 0pt]
      \lstinputlisting[language=bash, numbers=none]{compile.sh}
  \end{tcolorbox}

The idea of compiler-based randomization is that different users
generate and run different program variants, which complicates
potential code-reuse attacks.
MCR may also be used against timing side-channel attacks by inserting
random memory \texttt{load} instructions, however, in this context,
the tool introduces a very high execution-time overhead of up to eight
times slowdown~\cite{crane}.

The compilation time of the tool is comparable with LLVM.
Similarly, the execution-time overhead is typically small (and
controllable), up to 25\% for no-operation insertion~\cite{multicomp},
however, the execution-time overhead is high when mitigating timing
side-channel attacks~\cite{crane}.

In summary, MCR is appropriate for defending against code-reuse
attacks, but it is not highly efficient as a mitigation of timing
side-channel attacks.
An additional drawback is that it is based on a relatively old version
of LLVM (see Table~\ref{tab:tools}).
However, the tool is available online, has small compilation time, and
supports whole-program compilation.

\subsection{SecDivCon}

Secure-by-construction Code Diversification
(SecDivCon)~\cite{secdivcon} is a constraint-based compiler that
allows the generation of efficient and secure code.
SecDivCon is open source\footnote{SecDivCon:
  \url{https://github.com/romits800/secdivcon_experiments.git}}. It is
based on Unison, a combinatorial compiler
backend\footnote{Unison: \url{https://github.com/unison-code/unison}}.
The tool is compatible with LLVM version 3.8 and targets a generic
MIPS processor and ARM Cortex M0.

SecDivCon defines compiler-backend transformations and security
requirements as a set of constraints.
SecDivCon uses a constraint solver to find diversified code
implementations that are highly optimized with regard to execution
time and secure against timing side channels.
The tool enables constant-resource mitigations and backend
diversification, such as no-operation insertion, hardware register
diversification, instruction shuffling, and register and memory copy
insertion.

To mitigate timing side channels, SecDivCon requires defining the
security policy, namely which variables are security sensitive
(Secret) and which variables do not leak any information (Public).
For example, to ensure that the code in Figure~\ref{lst:cr} is
constant resource, the user need to define the following policy:
  \begin{tcolorbox}[colback=\codebackcolor, colframe=\codebackcolor,
        top=-8pt, bottom = -10pt, right = 0pt, left = 0pt]
\begin{lstlisting}[style=spstyle, numbers=none]
Public a0;
Secret a1
\end{lstlisting}
  \end{tcolorbox}
  Notations \lstinline[style=spstyle]{a0} and
  \lstinline[style=spstyle]{a1} correspond to the argument registers
  that contain the input arguments.
Given the security policy, SecDivCon generates automatically diverse
programs that are secure against timing side channels.

SecDivCon has several advantages: 1) combines performance and security
goals, 2) guarantees the preservation of properties in the generated
code, and 3) does not require extensive annotations.
These advantages come at the cost of compilation complexity, which
makes SecDivCon practical for small, vulnerable cryptographic
functions.
Thus, SecDivCon needs to be combined with different compilation
approaches to compile large code bases.

In summary, SecDivCon offers strong guarantees for mitigating timing
side channels for predictable devices, and provides a fine-grained
diversification scheme, with high gadget diversification ability.
The main disadvantage of SecDivCon is the high compilation time, the
lack of support for advanced computer architectures, and the
dependence on a relatively old version of LLVM.
\vspace*{-8pt}

\subsection{Partial Control-Flow Linearization (PCFL)}

Partial Control-Flow Linearization (PCFL)~\cite{linearization} is
a linearization tool that converts a non-constant-time program to
constant time.
PCFL is available online\footnote{PCFL:
  \url{https://github.com/lac-dcc/lif}} and uses LLVM 13.

The tool requires annotation of secret and non-secret values and is
able to generate constant-time programs automatically without
requiring manual effort from the developer.
The following code snippet, shows how to use PCFL to mitigate the
function \lstinline[style=cstyle]{check_bit} in Figure~\ref{lst:cr}.

\begin{tcolorbox}[colback=\codebackcolor, colframe=\codebackcolor,
        top=-8pt, bottom = -10pt, right = -5pt, left = -8pt]
 \lstinputlisting[style=cstyle, numbers = none, showlines=true,
   basicstyle = \ttfamily\small]{pcfl.c}
\end{tcolorbox}

Notation \lstinline[style=cstyle]{secret} ensures that the compiler
generates constant-time code for the annotated memory locations.
Note that PCFL is able to handle more complex programs, for example,
\lstinline[style=cstyle]{for} loops.

Table~\ref{tab:tools} shows a high execution-time overhead for the
transformed code by PCFL, however, the comparison considers non
constant-time code, which is not the case with the work by \citet{vu},
where the source code is already constant time.
PCFL focuses on timing side channels and is able to provide guarantees
in the generated code.
This approach has several advantages including, automatic
constant-time code generation, low compilation overhead, and high
program coverage.

Similar to the work by \citet{vu}, PCFL does not support mitigations
against code-reuse attacks apart from the LLVM optional compiler
mitigations.
Performing advanced post-compilation mitigations would require
reassessing the mitigation guarantees by PCFL.
In summary, PCFL is a potential solution for reducing the overhead and
final-code uncertainty of manual constant-time implementations.
\vspace*{-8pt}

\section{OPEN CHALLENGES}

The different compilation approaches we discussed in the previous
section provide possible solutions for compiling cryptographic
libraries.
These tools are based on LLVM, which is mostly compatible with the
current compilation schemes of the cryptographic libraries in
Table~\ref{tab:tools}.

The work by \citet{vu} is among the most appropriate approaches for
dealing with timing side-channels because it provides control over the
compilation process and supports constant-time programming.
Unfortunately, the tool is not publicly available.
PCFL may be used in cryptographic libraries that do not provide any
constant-time source-code mitigations to protect potentially
vulnerable parts of the code automatically.
Similarly, SecDivCon may provide a method to mitigate timing attacks
in insecure source code for embedded devices.
SecDivCon may also be used in a diversification scheme to harden
cryptographic libraries against code-reuse attacks.
MCR provides whole-program diversification and may be used against
both code-reuse attacks and to hide timing vulnerabilities in
libraries that do not implement source-code timing mitigations.
However, MCR introduces significant overhead when used against timing
side channels, which may be an important hinder in adapting MCR in
performance-sensitive applications.
To summarize, depending on the focus of a cryptographic library and
the current implementation, different compilation approaches may be
valuable to enhance the security of the library.

Each of the proposed secure compilation approaches has limitations:
MCR only focuses on diversification, SecDivCon can compile only small
functions, the tool by \citet{vu} and PCFL do not target code-reuse
attacks, while the former is not available online.
In addition, none of the presented tools targets RISC-V, a popular
instruction-set architecture that is supported by LLVM.

We believe that bridging the optimization/security gap requires
additional effort:
\begin{itemize}
\item \textbf{General-purpose compiler developers:} need to provide
  more control over the compiler outcome, allowing control over
  optimizations.
\item \textbf{General-purpose compiler developers:} need to implement
  infrastructure and further security mitigations, including
  constant-time preservation.
\item \textbf{Compiler developers:} need to focus on generating
  optimized code by redesigning compiler optimizations to be security
  aware.
\item \textbf{Secure-compiler developers:} need to design tools that
  combine multiple mitigations instead of focusing on a single
  mitigation.
\item \textbf{Cryptographic-library developers:} should make sure to
  implement constant-time alternatives and enable security flags in
  the compiler, which we have seen to some extent in this article.
\end{itemize}
\vspace*{-8pt}

\section{CONCLUSION}

This article describes security mitigations in popular cryptographic
libraries and proposes the use of secure compilation approaches to
enhance their security.
More specifically, most popular libraries depend on manual mitigations
and general-purpose compilers that provide secure solutions to a
number of vulnerabilities.
Unfortunately, general-purpose compiler mitigations are not sufficient
against advanced attacks and do not always preserve source-code
mitigations.
Improving the security of cryptographic libraries requires changes in
1) the current development process of cryptographic libraries to
consider using secure compilers, 2) the compilation approaches to
allow for additional mitigations and more transparency, and 3) the
secure-compilation approaches to combine multiple mitigations in one
tool.

\vspace*{-8pt}

\section{ACKNOWLEDGMENTS}
The work of P. Papadimitratos is supported in part by the Knut and
Alice Wallenberg Academy Fellow Trustworthy IoT project.
E. Troubitsyna is partially supported by Swedish Foundation for
Strategic Research with project FUS21-0026 \textit{SUCCESS:
  Sustainable Cyber-Physical Software-Defined System Slicing}.

The authors would like to thank Peter Gutmann for his valuable input
on the interpretation of hardening check results and the role of
PIE/PIC in static and dynamic library builds.

\def\refname{REFERENCES}

\bibliographystyle{IEEEtranN}
\bibliography{bibliography}

\begin{IEEEbiography}{Rodothea Myrsini Tsoupidi}{\,}
  is a independent Researcher in Stockholm, Sweden.
Her research interests include compiler optimization, software
diversification, language-based security, and side-channel
mitigations.
She received a Ph.D on Information and Communication Technology from
the school of Electrical Engineering and Computer Science, at Royal
Institute of Technology KTH, Sweden.
\vspace*{8pt}
\end{IEEEbiography}

\begin{IEEEbiography}
  {Elena Troubitsyna}{\,} is a Professor at KTH -- Royal Institute of
  Technology, Department of Computer Science, where she leads a
  research group that studies techniques for the development of
  dependeable-by-construction systems.
  Her research focuses on creating formal and model-driven methods for
  the development of safety- and security-critical software-intensive
  systems.
  Elena Troubitsyna received her Ph.D. Degree from the Turku Center
  for Computer Science.
\vspace*{8pt}
\end{IEEEbiography}

\begin{IEEEbiography}
  {Panos Papadimitratos}{\,} leads the Networked Systems Security
  group and he is a member of the steering committee of the Security
  Link center, KTH.
  Papadimitratos received his Ph.D. degree in Electrical and
  Computer Engineering from Cornell University.
  He is a fellow of the Young Academy of Europe, a Knut and Alice
  Wallenberg Academy Fellow, an IEEE Fellow, and an ACM Distinguished
  Member.
  He serves or served as: member (and currently chair) of the ACM
  Conference on Security and Privacy in Wireless and Mobile Networks
  (WiSec) steering committee; member of the Privacy Enhancing
  Technologies Symposium (PETS) Editorial and Advisory Boards and the
  International Conference on Cryptology and Network Security (CANS)
  Steering Committee; program chair for the ACM WiSec 2016,
  International Conference on Trust \& Trustworthy Computing (TRUST)
  2016 and CANS 2018 conferences; general chair for ACM WiSec 2018, PETS
  2019 and IEEE European Symposium on Security and Privacy 2019
  conferences; and Associate Editor of the IEEE Transactions on Mobile Computing, ACM/IEEE
  Transactions on Networking, Institute of Engineering and Technology
  Information Security and ACM Mobile Computing and Communications
  Review journals. His group webpage is:
  \url{https://www.eecs.kth.se/nss}.
\vspace*{8pt}
\end{IEEEbiography}

\end{document}